\documentclass[%
twocolumn,superscriptaddress,groupedaddress,
showpacs,preprintnumbers,
nofootinbib,nobibnotes,
amsmath,amssymb,prl,
floatfix,
showpacs]{revtex4}

\usepackage{graphicx}
\usepackage{mwe}
\usepackage{float}
\usepackage{dcolumn}
\usepackage{bm}
\begin{document}
\title{Hydrodynamic Phonon Transport Perpendicular to Diffuse-Gray Boundaries}

\author{Runqing Yang}
\address{Department of Mechanical Engineering, University of California, Santa Barbara, CA 93106, USA}
\affiliation{Department of Physics, University of California, Santa Barbara, CA 93106, USA}
\author{Shengying Yue}
\author{Bolin Liao}\email{bliao@ucsb.edu}
\affiliation{Department of Mechanical Engineering, University of California, Santa Barbara, CA 93106, USA}

\date{\today}
             
\begin{abstract}
In this paper, we examine the application of an ideal phonon-hydrodynamic material as the heat transfer medium between two diffuse-gray boundaries with a finite temperature difference. We use the integral-equation approach to solve a modified phonon Boltzmann transport equation with the displaced Bose-Einstein distribution as the equilibrium distribution between two boundaries perpendicular to the heat transfer direction. When the distance between the boundaries is smaller than the phonon normal scattering mean free path, our solution converges to the ballistic limit as expected. In the other limit, we find that, although the local thermal conductivity in the bulk of the hydrodynamic material approaches infinity, the thermal boundary resistance at the interfaces becomes dominant. Our study provides insights to both the steady-state thermal characterization of phonon-hydrodynamic materials and the practical application of phonon-hydrodynamic materials for thermal management.
\end{abstract}


\maketitle

\section{\label{sec:level1}Intoduction}
Phonons are major carriers of heat in semiconductors and insulators and the scattering between phonons is usually the main source of thermal resistance in single crystals of these materials\citep{Ziman1960}. Phonon-phonon scattering can be classified into two types: the momentum-conserving normal scattering processes and the momentum-destroying Umklapp scattering processes. It is understood\citep{Peierls1929} that Umklapp scatterings act as momentum sinks in the bulk of a material and directly contribute to the thermal resistance. Whereas normal scattering processes do not directly create thermal resistance \textit{per se}, they perturb the phonon distributions and indirectly affect the thermal transport in presence of Umklapp processes\citep{Callaway1959, maznev2014demystifying}. In different materials and under different external conditions, the dominating phonon-phonon scattering mechanism can vary, giving rise to distinct regimes of heat conduction\citep{Chen2005,Cepellotti2015,Liao2015MRS}. When the characteristic size of the sample is smaller than the intrinsic phonon mean free path due to phonon-phonon scatterings, as is relevant in microelectronic devices\citep{Schleeh2015} and nanostructured thermoelectric materials\citep{Poudel2008}, the extrinsic phonon-boundary scattering dominates and the phonon transport approaches the ballistic regime\citep{Maasilta2014,Minnich2011}. In macroscopic samples, where the intrinsic phonon-phonon scattering becomes the leading phonon scattering channel, the relative strength of the normal processes and Umklapp processes determines the characteristics of the phonon transport. In most three-dimensional bulk materials above their Debye temperature, the Umklapp processes prevail and the phonon transport is diffusive, where no net drift flow of phonons can be established and maintained due to the constant dissipation of the phonon momenta, leading to the familiar Fourier’s law of heat conduction. On the other hand, if the normal scattering becomes the much more frequent scattering process, the phonon momenta are largely conserved during the transport. In this regime, phonons can develop a nonzero drift velocity when they are subjected to a temperature gradient, analogous to a viscous fluid system driven by a pressure gradient. With this analogy, this regime of heat conduction is named the phonon hydrodynamic transport regime\citep{Gurevich1986,Guyer1966-1,Guyer1966-2}.
\begin{figure}
  \centering
   \includegraphics[width=84mm]{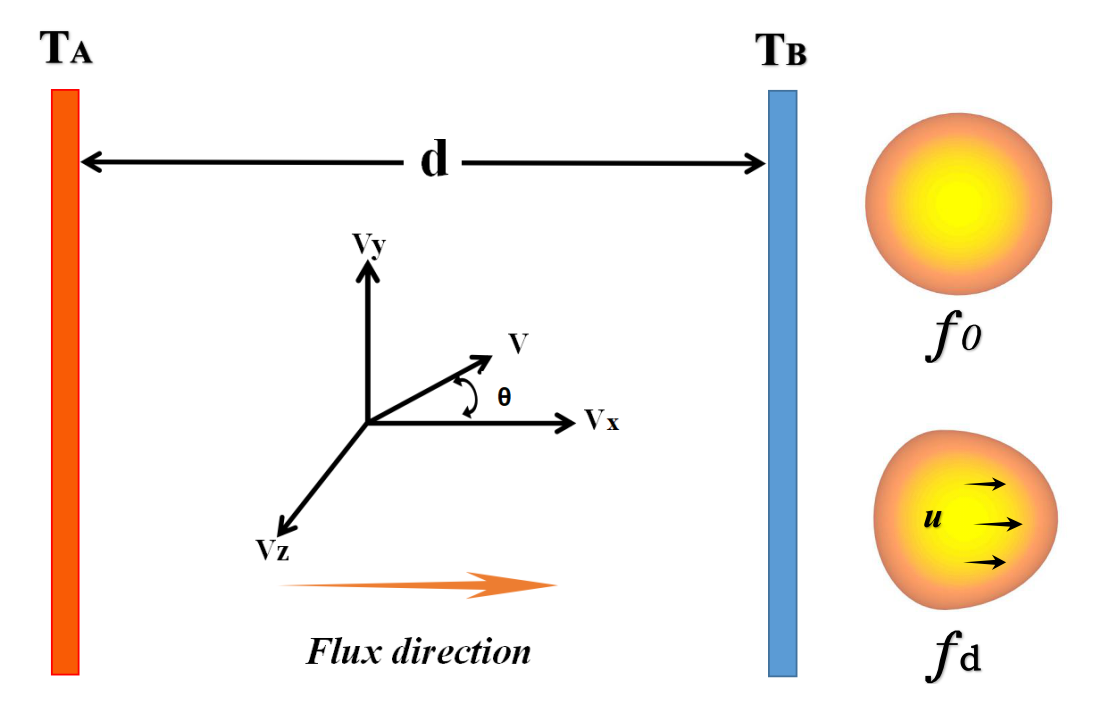}
  \caption{(Left) The geometry and the coordinate system used for the phonon heat transfer problem in this work. $T_A$ and $T_B$ are temperatures at the boundaries. (Right) Comparison of the normal Bose-Einstein distribution $f_0$ and the displaced Bose-EInstein distribution $f_d$. Two iso-frequency surfaces of the two distributions are plotted to demonstrate their difference in the symmetry with regard to the phonon wavevector directions.}
\end{figure}

Although the phonon hydrodynamic transport holds the promise of low dissipation and more efficient heat conduction and exhibits interesting transport phenomena such as the phonon Poiseuille flow and the second sound\citep{Guyer1966-1,Guyer1966-2}, its occurrence in three-dimensional bulk materials requires rather stringent conditions and has only been observed at cryogenic temperatures in solid helium\citep{Ackerman1966} and a few other materials\citep{Narayanamurti1972,Jackson1970,Pohl1976,Hehlen1995,Koreeda2007}. Recently, the research interests in phonon hydrodynamic transport have been revived thanks to the first-principles-simulation-based predictions of phonon hydrodynamic transport happening in low-dimensional\citep{Lee2015,Cepellotti2015,Lee2017} and layered materials\citep{Ding2017} at more practical temperatures, due to the highly anharmonic zone-center flexural phonon modes\citep{Lee2015}. This promising development indicates a great potential of exploiting hydrodynamic phonon transport in these materials for practical thermal management applications. For experimental verification and practical applications, however, it is important to consider hydrodynamic phonon transport constrained by realistic geometry and boundary conditions. Li and Lee analyzed hydrodynamic phonon transport in suspended graphene with diffuse boundaries parallel to the heat flux direction using a Monte-Carlo simulation\citep{Li2018}. Guo and Wang analyzed the same geometry using a discrete ordinate method to solve the phonon Boltzmann transport equation\citep{Guo2017}. 

In this paper, we present a method to analyze the hydrodynamic phonon transport with two diffuse-gray boundaries perpendicular to the heat transfer direction, as illustrated in Fig.1. This scenario arises if a phonon-hydrodynamic material is used as a heat transfer medium between two normal materials, or if a steady-state thermal measurement is performed on a phonon-hydrodynamic material with two normal contacts. Cepellotti and Marzari\citep{cepellotti2017boltzmann} analyzed heat transport in MoS\textsubscript{2} in the same geometry based on first-principles calculation and the friction process of relaxons. Our method is based on the integral solution of the phonon Boltzmann equation widely adopted to analyze cross-plane ballistic phonon transport in thin films\citep{Chen1998}. To obtain an analytical framework that helps clarify the essential physical features of phonon hydrodynamic transport, we consider an ideal material with only normal phonon scattering processes in this work. We present our method and analytical results in Section II and then analyze our results in Section III by carefully comparing them to the non-hydrodynamic case at different acoustic thicknesses. Our results could improve the understanding of the phonon hydrodynamic transport and lead to further considerations of practical applications of phonon-hydrodynamic materials.  

\section{\label{sec:method}Method}
In the hydrodynamic phonon transport regime, the normal processes dominate and will drive phonons to equilibrate toward
a displaced Bose-Einstein distribution \citep{Guyer1966-1} which can be written as
\begin{equation}
f_{d}=\frac{1}{exp[\frac{\hbar(\omega-\textbf{k} \cdot \textbf{u})}{k_{B}T}]-1},
\end{equation}
where $\hbar$,$\omega$,$\mathbf{k}$,$k_B$, and T denote the reduced Planck constant,
phonon frequency, phonon wavevector, the Boltzmann constant,
and temperature, respectively. The drift velocity, \textbf{u}, is constant at a given location for all phonon modes. A comparison of the normal and the displaced Bose-Einstein distribution is given in Fig. 1. One distinct feature of the displaced Bose-Einstein distribution is its asymmetry with regard to phonon traveling directions, implying that a finite heat flow can occur even when phonons approach their local equilibrium through frequent normal phonon-phonon scatterings. Assuming a small temperature gradient and drift velocity, (1)
can be linearized \citep{Ding2017} to 
\begin{equation}
f_{d}=f_{0}+f_{0}(f_{0}+1)\frac{\hbar}{k_{B}T}\textbf{k} \cdot \textbf{u},
\end{equation}
where $f_0$ is the normal Bose-Einstein distribution.

The phonon Boltzmann transport equation is commonly used to solve for the distribution functions of phonons in nonequilibrium transport scenarios. Within the relaxation time approximation, the complex phonon scattering terms are replaced with a phenomenological term that describes the equilibration effect of phonon scattering events: the Umklapp scatterings equilibrate phonons towards the Bose-Einstein distribution, whereas the normal scatterings equilibrate phonons towards the displaced Bose-Einstein distribution. When both scattering mechanisms are present, both terms are included in the phonon Boltzmann transport equation, which is often referred to as Callaway's dual relaxation model\citep{Callaway1959}.  We note that quantitative predictions of the thermal conductivity of real materials can be made by solving the phonon Boltzmann transport equation with the full scattering matrix iteratively\citep{broido2007intrinsic} or with other numerical methods\citep{guo2016lattice,zhao2009phonon,peraud2011efficient,guo2015phonon}, whereas the relaxation time approximation often has the advantage of providing analytical results, which are more convenient for theoretical analyses. 

Here we consider an ideal material with only normal phonon scattering processes and the corresponding phonon Boltzmann transport equation takes the form:
\begin{equation}
\frac{\partial f}{\partial t} + \textbf{v}_g \cdot \nabla f=- \frac{f-f_{d}}{\tau},  
\end{equation}
where $f$ is the phonon distribution function, $f_d$
the equilibrium distribution (the displaced Bose-Einstein distribution), $\mathbf{v}_{g}$ the phonon group
velocity, and $\tau$ the phonon relaxation time due to normal scattering processes. For one-dimensional steady state situations, equation(3) becomes 
\begin{equation}
{v}_g \cos{\theta} \frac{\partial f}{\partial x}=- \frac{f-f_{d}}{\tau},
\end{equation} 
where $\theta$ is a polar angle between a phonon wavevector direction and the heat transfer direction along the positive-x axis. We further assume the two boundaries perpendicular to the heat transfer direction are diffuse and gray and do not support hydrodynamic phonon transport, namely phonons entering and leaving the hydrodynamic material assume the normal Bose-Einstein distribution. A complete treatment of the interface between a hydrodynamic material and a non-hydrodynamic material would require solving the phonon Boltzmann transport equation in both materials and the transmission of different phonon modes at the interface, which is beyond the scope of this work. Nevertheless, our model can capture the transition of the phonon distribution function between two different local equilibrium distributions at the interfaces and examine the additional thermal boundary resistance introduced by this transition. 

The solution of Eq.(4) has the following integral form:
\begin{equation}
f_+=\frac{1}{e^{\frac{\hbar\omega}{k_BT_A}}-1}e^{-\frac{m}{\mu}}+\int_0^{m}e^{-\frac{m-m'}{\mu}}f_d \frac{dm'}{\mu} \ \ (\mu > 0),
\end{equation}
\begin{equation}
f_-=\frac{1}{e^{\frac{\hbar\omega}{k_BT_B}}-1}e^{\frac{\xi - m}{\mu}}-\int_{m}^{\xi}e^{-\frac{m-m'}{\mu}}f_d \frac{dm'}{\mu} \ \ (\mu < 0),
\end{equation}
where $m=\frac{x}{\Lambda}$ ($\Lambda=v_g \tau$ is the phonon mean free path due to normal scattering processes), $\xi=\frac{d}{\Lambda}$ (the acoustic thickness, with $d$ being the thickness of the sample), and $\mu=\cos\theta$. Since both the local temperature and the local drift velocity of the phonons are unknown, two additional conditions are required to close the problem. In phonon hydrodynamic transport, as both energy and momentum of the phonons are conserved \citep{Chen2005,Cepellotti2015}, the energy flux $\textbf{$J$}$ and the momentum flux $\textbf{$Q$}$ should remain constant along the transport direction at the steady state. Due to the symmetry, physical quantities along y and z directions are naturally conserved, so we only need to consider the energy and momentum fluxes along the x direction. Such a model is analogous to the one-dimensional parallel-plate participating medium model in radiative heat transfer\citep{M}. Using the same reasoning, we conclude that the drift velocity is also along the x direction. Then the conservation conditions of energy and momentum of phonons, assuming three degenerate acoustic branches, lead to the following relations:
\begin{equation}
 \frac{\partial \textbf{$J_x$}(m)}{\partial m}=0, \textmd{where} \ \textbf{$J_x$}(m)=\int v_{g}\mu \hbar \omega_k f(m,k) \frac{3\mathbf{d^{3}k}}{(2\pi)^{3}},
\end{equation}

\begin{equation}
 \frac{\partial \textbf{$Q_x$}(m)}{\partial m}=0, \textmd{where} \ \textbf{$Q_x$}(m)=\int v_g\mu \hbar k_x f(m,k) \frac{3\mathbf{d^{3}k}}{(2\pi)^{3}}.
\end{equation}
Eqs.(7) and (8) also serve as the definitions of effective temperature and drift velocity in this non-equilibrium case\citep{casas2003temperature}.  

Since the phonon hydrodynamic transport is predicted to happen at temperatures much below the Debye temperature, we use the Debye model  $w=v_g|\mathbf{k}|$ as the dispersion relation of acoustic phonons and assume the wavevector $\mathbf{k}$ can be integrated to infinity. The conservation laws (7)(8) provide all constraints required to derive the temperature and the drift velocity distributions in our calculation. Replacing $f_d$ with the expansion in Eq. (2), Eqs. (5) and (6) provide the solution of the nonequilibrium phonon distribution function $f$ inside the material, with the temperature $T$ and the phonon drift velocity $\mathbf{u}$ undetermined. Substituting Eqs. (5) and (6) into Eqs. (7) and (8) and carrying out the integration of the phonon momentum $\mathbf{k}$ lead to the following equations, from which the distributions of $T$ and $\mathbf{u}$ can be solved:
\begin{widetext}
\begin{equation}
\begin{split}
J_{q1}^+ E_2(m)+J_{q2}^-E_2(\xi-m)=&2e_0(m)- \int_{0}^{m} e_0(m^{'}) E_1(m-m^{'}) dm'- \int_{m}^{\xi} e_0(m^{'}) E_1(m^{'}-m) dm'\\
&- \int_{0}^{m} \frac{4u(m{'})}{v_g} e_0(m^{'}) E_2(m-m^{'}) dm'+\int_{m}^{\xi} \frac{4u(m{'})}{v_g} e_0(m^{'}) E_2(m^{'}-m) dm',
\end{split}
\end{equation}
\begin{equation}
\begin{split}
J_{q1}^{+} E_3(m)-J_{q2}^- E_3(\xi-m)  = &- \int_{0}^{m} e_0(m^{'}) E_2(m-m^{'}) dm' + \int_{m}^{\xi} e_0(m^{'}) E_2(m^{'}-m) dm'\\
&+\frac{8u(m)}{3v_g}e_0(m)-\int_{0}^{m} \frac{4u(m{'})}{v_g}e_0(m^{'}) E_3(m-m^{'}) dm'
-\int_{m}^{\xi} \frac{4u(m{'})}{v_g} e_0(m^{'}) E_3(m^{'}-m) dm',
\end{split}
\end{equation}
\end{widetext}
where,
$J_{q1^+}=\sigma {T_A}^4$ ($\sigma= \frac{\pi^2 k_B^4}{20 \hbar^3 v_g^2}$ is a constant, whose form is analogous to the Stefan-Boltzmann constant in photon transport \citep{Chen2005}),
$J_{q2^-}=\sigma {T_B}^4$, $E_n(x)= \int_0^1 \mu^{n-2}e^{-\frac{x}{\mu}}d\mu$,
$e_0(m)=\sigma {T(m)}^4$. $e_0(m)$ is the local phonon emissive power and reflects the local phonon temperature. $T_A$ and $T_B$ are the boundary temperatures; $J_{q1^+}$ and $J_{q2^-}$ are the phonon heat fluxes leaving the left boundary (A) and entering the right boundary (B), respectively. Equations (9) and (10) can be further simplified: by differentiating both sides of (10) with respect to $m$, we have:
\begin{widetext}
\begin{equation}
\begin{split}
-\bigg[J_{q1}^+ E_2(m)+J_{q2}^- E_2(\xi-m)\bigg]=
&-\bigg[2e_0(m)- \int_{0}^{m} e_0(m^{'}) E_1(m-m^{'}) dm'-\int_{m}^{\xi} e_0(m^{'}) E_1(m^{'}-m) dm'\\
&- \int_{0}^{m} \frac{4u(m{'})}{v_g}e_0(m^{'}) E_2(m-m^{'}) dm'+\int_{m}^{\xi} \frac{4u(m{'})}{v_g} e_0(m^{'}) E_2(m^{'}-m) dm'\bigg]  \\ 
&+\bigg[\frac{8u(m)}{3v_g}e_0(m)\bigg]^{'}.
\end{split}
\end{equation}.
\end{widetext}
By comparing (11) with (9), it is shown that the product of the local emissive power and the drift velocity is a constant along the x direction, so we can denote $\frac{4u(m)}{v_g}e_0(m)$ as a constant Y. As a result, Eqs.(9)(10) can be simplified to the following integral equation set:
\begin{widetext}
\begin{equation}
J_{q1}^+ E_2(m)+J_{q2}^-E_2(\xi-m)=2e_0(m)- \int_{0}^{m} e_0(m^{'}) E_1(m-m^{'}) dm'- \int_{m}^{\xi} e_0(m^{'}) E_1(m^{'}-m) dm'+[E_3(m)-E_3(\xi-m)]Y,
\end{equation}
\begin{equation}
J_{q1}^{+} E_3(m)-J_{q2}^- E_3(\xi-m)  = - \int_{0}^{m} e_0(m^{'}) E_2(m-m^{'}) dm' + \int_{m}^{\xi} e_0(m^{'}) E_2(m^{'}-m) dm'+[E_4(m)+E_4(\xi-m)]Y.
\end{equation}
\end{widetext}

Equations (12) and (13) can be converted into matrix equations by discretizing the spatial coordinate $m$. The unknowns are $e_0(m)$, a vector, and $Y$, a scalar, and the integration kernels $E_n$ can be converted into matrices. The resulted equations can be solved numerically by matrix inversion\citep{Chen2005}.
\section{\label{sec:Results and discussion}Results and Discussion}
\begin{figure*}
 \includegraphics[width=0.7\linewidth,clip]{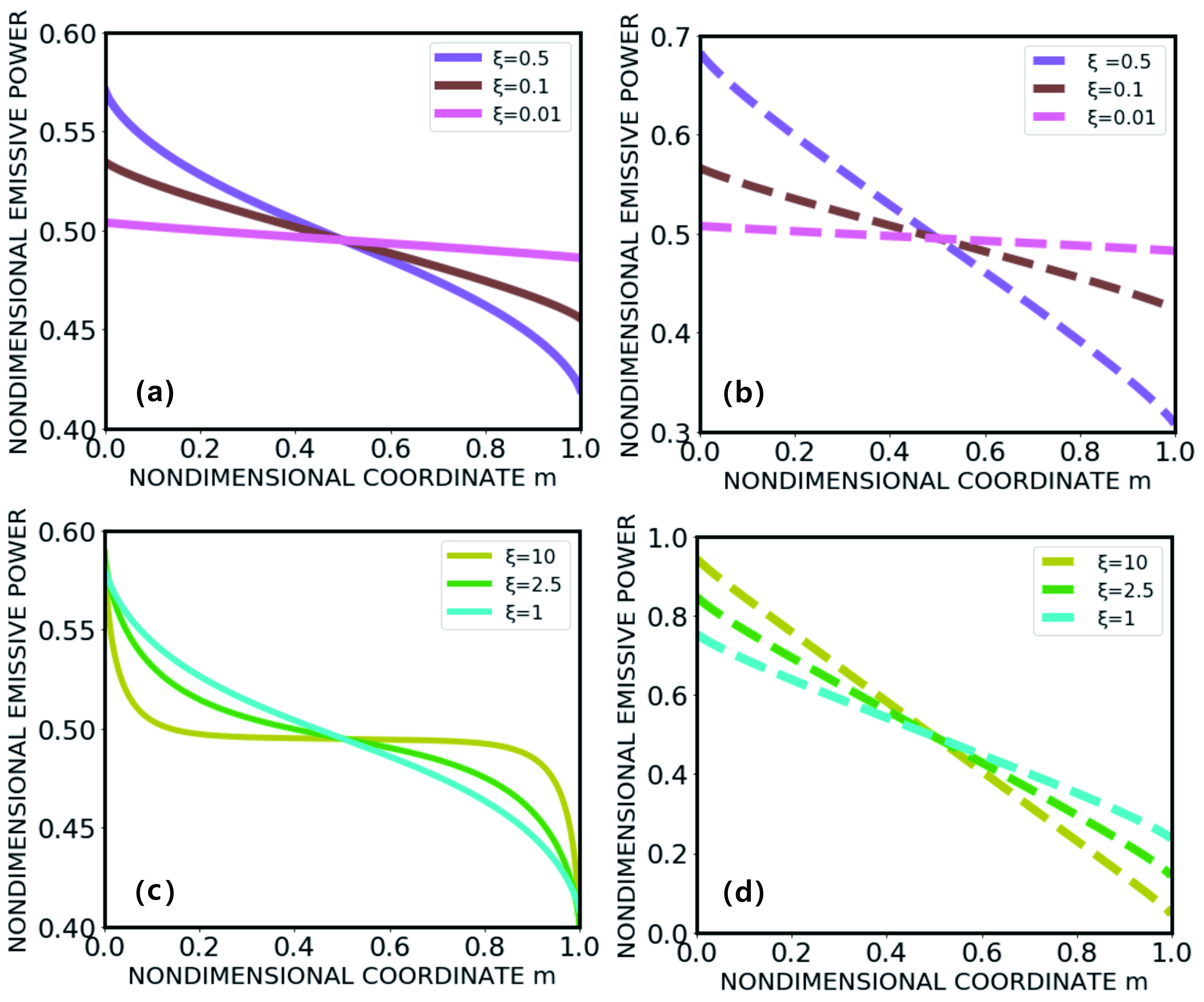}
  \caption{The distribution of the nondimensional phonon emissive power $J^{*}(m)=\frac{e_0(m)-J_{q2}^-}{J_{q1}^{+}-J_{q2}^-}$ in the hydrodynamic and non-hydrodynamic regimes at different acoustic thicknesses $\xi$: solid lines represent the normalized phonon emissive power in the hydrodynamic regime and the dotted lines represent those in the non-hydrodynamic regime. The same-color lines denote the same acoustic thickness.}
\end{figure*}
By definition, if Umklapp scatterings dominate in the system, we would not have the momentum conservation equation (13), and Eq.(12) would reduce to the energy conservation equation in non-hydrodynamic systems as derived previously\citep{Chen2005}. To contrast the differences between the hydrodynamic and non-hydrodynamic phonon transport, we first compare the nondimensional local phonon emissive power (a measure of the temperature profile) $J^{*}(m)=\frac{e_0(m)-J_{q2}^-}{J_{q1}^{+}-J_{q2}^-}$ in both cases at two limits: $\xi\ll1$ and $\xi\gg1$, as shown in Fig. 2.

When the phonon mean free path is very large ($\xi\ll1$), the hydrodynamic transport shows similar behavior as the non-hydrodynamic case due to the lack of scattering in both cases, as shown in Fig.2(a) and (b). In this ballistic limit, phonons do not lose energy or momentum during the transport inside the material due to phonon-phonon scatterings. Like transport in the non-hydrodynamic case, the hydrodynamic phonon system has temperature jumps at the boundaries due to thermalization of the nonequilibrium phonon distributions near the boundaries. 
\begin{figure}
\includegraphics[width=1.0\linewidth,clip]{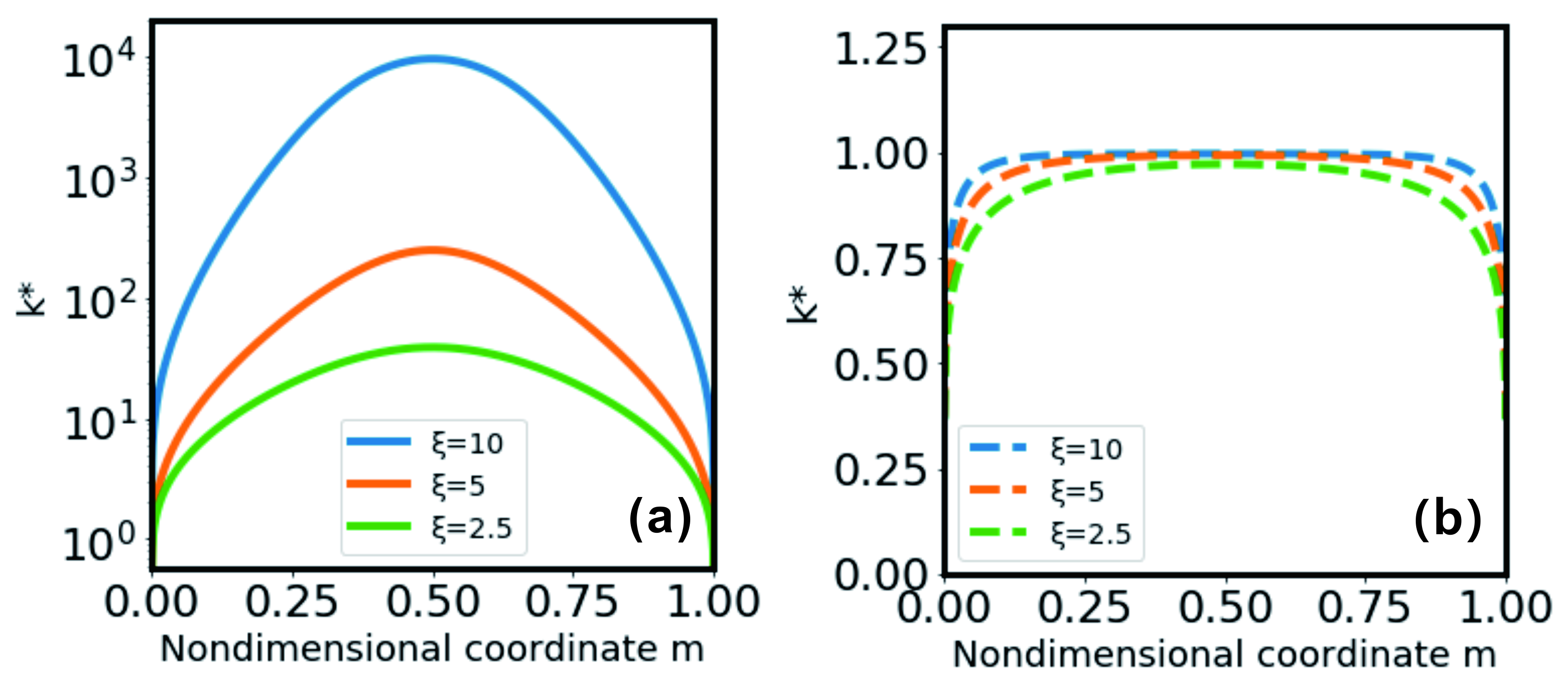}
\caption {The local effective thermal conductivity $k^*$ in hydrodynamic and non-hydrodynamic cases at different acoustic thickness $\xi$: the solid lines represent the non-dimensional phonon emissive power in the hydrodynamic case and the dotted lines represent the non-dimensional phonon emissive power in the non-hydrodynamic case. Lines with the same color denote the same acoustic thickness.}
\end{figure}
On the other hand, when the phonon mean free path is small($\xi\gg1$) compared to the sample thickness, the phonon emissive power in the hydrodynamic case is highly nonlinear and shows completely different behavior compared to the non-hydrodynamic case. In this ``continuum limit"\citep{BULUSU,continuum1}, phonon scatterings are frequent and drive the local phonon distributions towards the corresponding equilibrium distributions. In the non-hydrodynamic case, the local phonon distribution in the bulk approaches the normal Bose-Einstein distribution, except for a small deviation proportional to the local temperature gradient\citep{Chen2005} (the normal Bose-Einstein distribution is symmetric in terms of phonon wavevector direction and cannot support a finite heat flux). This local phonon distribution gives rise to the Fourier's law: the temperature distribution in the bulk approaches linear and the temperature jumps at boundaries gradually disappear as $\xi$ increases. In the hydrodynamic transport regime, however, the frequent normal scatterings will drive the phonons towards the displaced Bose-Einstein distribution, which can support a finite heat flux due to its asymmetry with regard to the phonon wavevector directions. Assuming the displaced Bose-Einstein distribution in the bulk, one can derive, instead of the Fourier's law, two continuum equations from the energy and momentum conservation conditions for the temperature and drift velocity distributions\citep{Chen2005,Gurevich1986}\\
\begin{align}
\frac{\partial(\eta_{ij})u_j}{\partial t}+\frac{\partial T}{\partial x_i}&=0\\
\frac{\partial U}{\partial t}+S_p T\frac{\partial u_i}{\partial x_i}&=0
\end{align}
Here, $\eta_{ij}$ is a constant second order tensor, $U$ is the total phonon energy density and $S_p$ is the phonon entropy density. The equation set leads to constant T and u in a 1D steady-state problem in the bulk, indicating an infinite bulk thermal conductivity as a finite heat flux flows isothermally. This is consistent with the physical picture that the thermal conductivity should approach infinity if only normal processes are present, due to the absence of dissipation \citep{Peierls1929,Ziman1960}. Our solution of the phonon Boltzmann transport equation in this regime captures this transition, in contrast to the non-hydrodynamic case. As shown in Fig. 2(c), as $\xi$ increases in the hydrodynamic regime, the phonon emissive power distribution in the bulk gradually approaches a constant, while the temperature jumps increase at the boundaries, reflecting the incompatibility of the boundary conditions with the continuum equations (14)(15) and the existence of a thermal boundary resistance at the hydrodynamic/non-hydrodynamic interfaces. This is a major difference between the hydrodynamic and non-hydrodynamic transport regimes. Sussmann and Thellung obtained a qualitatively similar distribution of the effective phonon temperature by deriving a set of differential equations for $T$ and $u$ assuming that the local equilibrium condition is valid\citep{sussmann1963thermal}. In comparison, our results are more general in that the local equilibrium condition is not required. They also imposed explicit boundary conditions for $T$ and $u$ so that the discontinuities of $T$ and $u$ at the boundaries were not captured\citep{sussmann1963thermal}. 

To further illustrate this difference, we can artificially define a local effective thermal conductivity $k_{\textrm{eff}}$ by imposing the Fourier's law for both hydrodynamic and non-hydrodynamic regimes, as shown in Fig. 3. 
Here, since $e_0= v_g \int_0^{\infty} \hbar \omega f(\omega) D(\omega)d\omega \approx v_g C \Delta T$, where $C$ is the volumetric heat capacity, we can define $k_{\textrm{eff}}$ as follows:
\begin{align}
J_x&=-k_{\textrm{eff}} \frac{d T}{d x}=-k_{\textrm{eff}} {(\frac{de_0}{dT})}^{-1} \frac{de_0}{dx}=-k_{\textrm{eff}}{(v_g C \Lambda)}^{-1} \frac{de_0}{dm}. 
\end{align}
To compare $k_{\textrm{eff}}$ and the kinetic formula of the bulk thermal conductivity $k_{\textrm{bulk}}=\frac{1}{3}Cv_g\Lambda$, it is convenient to calculate the ratio of the two quantities $k^*=-\frac{3J_x}{(\frac{\partial e_0}{\partial m})}$.
From Fig. 3(b), we can see that at the continuum limit ($\xi$ is large), $k^*$ approaches 1 in the non-hydrodynamic case, as expected. In contrast, in the hydrodynamic case, there is no upper bound to $k^*$ in the bulk as $\xi$ increases, whereas finite thermal resistances exist at the boundaries.\\ 
\begin{figure}
  \centering
    \includegraphics[width=0.8\linewidth,clip]{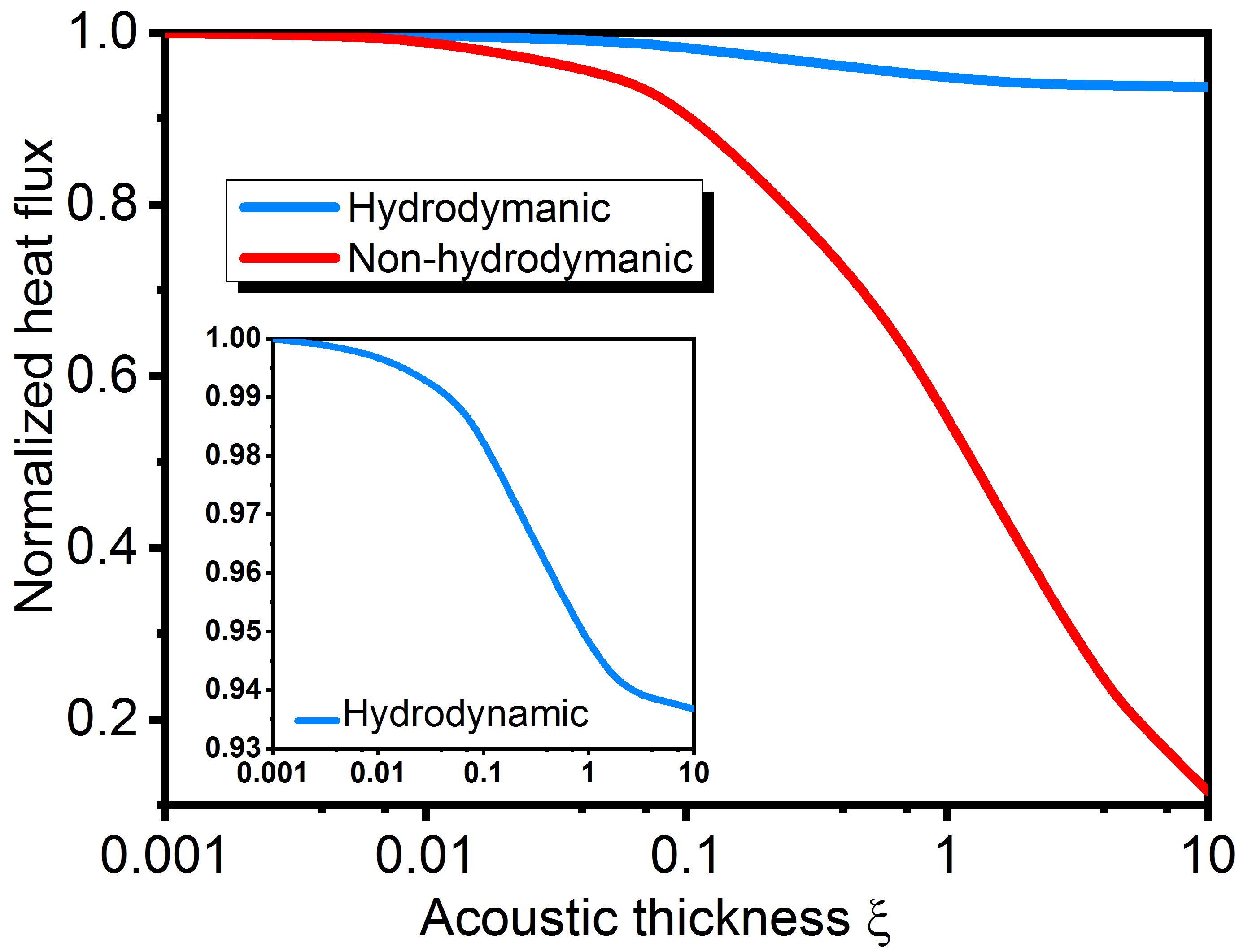}
  \caption{The transported heat flux in the hydrodynamic (blue line) and non-hydrodynamic (red line) regimes with fixed boundary temperatures, normalized to the value at the ballistic limit.}
\end{figure}
We further compare the heat fluxes transported across the medium under the same temperature difference in hydrodynamic and non-hydrodynamic cases. The heat flux as a function of the acoustic thickness is shown in Fig. 4.  When $\xi$ is small, the heat fluxes in both hydrodynamic and non-hydrodynamic cases approach the ballistic limit. As $\xi$ increases, the heat flux in the non-hydrodynamic case decreases due to the increased Umklapp scatterings and dissipation, and eventually approaches zero as the temperature gradient approaches zero. In the hydrodynamic case, however, the heat flux reduces from the ballistic limit, remains higher than that in the hydrodynamic case and approaches a finite value as $\xi$ approaches infinity. Since the bulk thermal conductivity approaches infinity in this case, the reduced heat flux from the ballistic limit is due to the boundary thermal resistances. The different heat fluxes in hydrodynamic and non-hydrodynamic cases confirm the superior heat transfer capability of phonon-hydrodynamic materials, even when sandwiched between two non-hydrodynamic materials. When both Umklapp and normal scattering processes are present, the heat flux transported will be in between the two bounds given in Fig. 4. 

Another important finding from our derivation is that the product of the local phonon emissive power and the local phonon drift velocity is a constant across the space, denoted by Y, given each $\xi$. Constant Y is the result of combining both energy and momentum conservation conditions and it relates the local phonon drift velocity and the phonon emissive power in a simple way. With the value of Y and the phonon emissive power we obtain from the numerical solution of (12) and (13), the change of the phonon drift velocity across the medium $\Delta u$ as a function of the temperature difference between the two boundaries can be calculated, as shown in Fig. 5(a) for different $\xi$ values. From Fig. 5(a), it is seen that, given the same acoustic thickness, the change of the phonon drift velocity increases rapidly as the difference between the phonon emissive power (and thus temperature) at the two boundaries increases, which is the driving force for the hydrodynamic transport. 
\begin{figure}
  \centering
 \includegraphics[width=1.0\linewidth,clip]{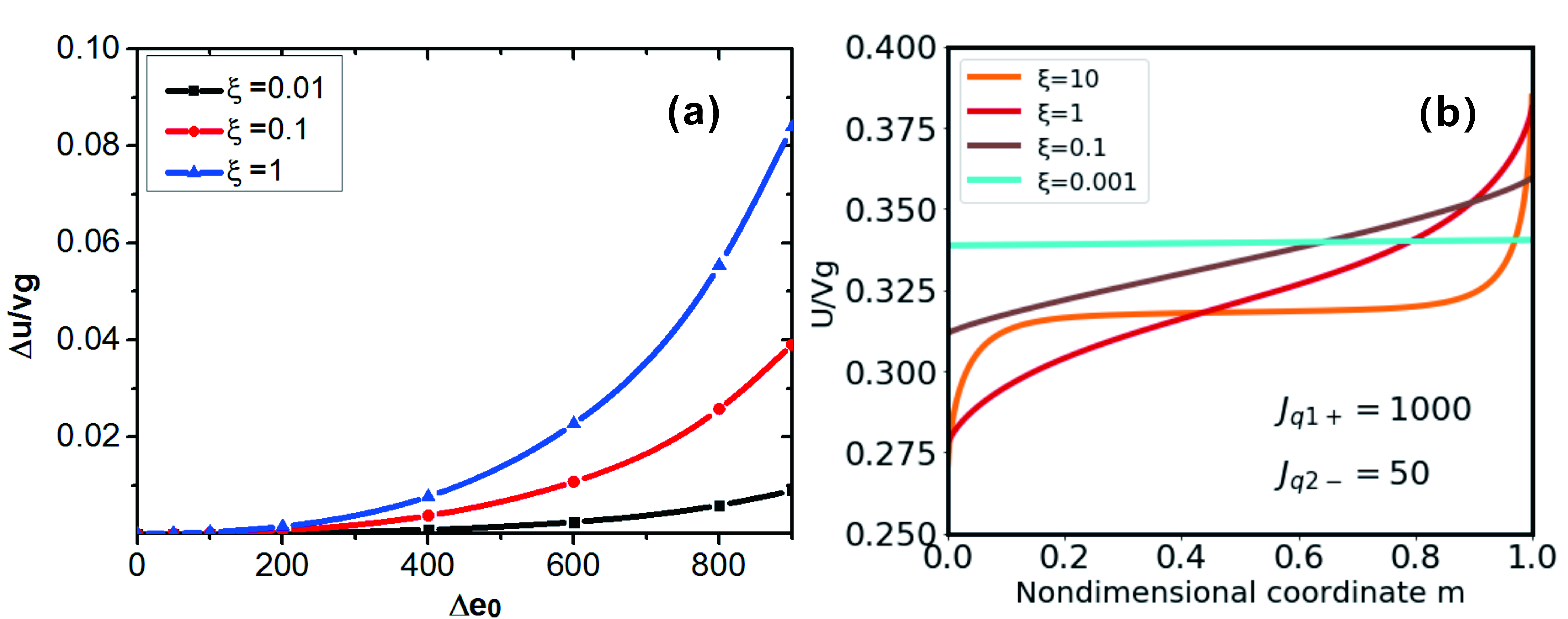}
  \caption{(a) The increase of the phonon drift velocity versus the phonon emissive power difference between the two boundaries at different acoustic thicknesses; (b) the spatial distribution of the phonon drift velocity at different acoustic thicknesses.}
\end{figure}

Since the product of the phonon emissive power and the local drift velocity is a constant across the space, the spatial distribution of the local drift velocity is simply the reciprocal of the distribution of the phonon emissive power, as shown in Fig. 5(b) (normalized to the group velocity $v_g$). We compare the u distributions at two limits: $\xi\ll1$ and $\xi\gg1$. When $\xi\ll1$, the lack of scattering makes the emissive power approach a constant; Y, as mentioned previously, is always a constant at any $\xi$, and thus u is almost constant across the domain. When $\xi\gg1$, the drift velocity distribution approaches a constant in the bulk, while changing rapidly near the boundaries, corresponding to the reciprocal of the phonon emissive power distribution. 

In summary, we have applied an integral-equation approach to solve the phonon Boltzmann transport equation with only normal scattering processes. With this approach, we have calculated the spatial distributions of the phonon emissive power and the drift velocity in a phonon-hydrodynamic material when the transport is perpendicular to two diffuse-gray boundaries. Our results reveal distinct features and offer mechanistic understanding of phonon hydrodynamic transport by comparing it with the non-hydrodynamic case, particularly at the continuum limit.

\section{Acknowledgement}
This work is supported by a startup fund from University of California, Santa Barbara (UCSB). B. L. acknowledges the support from a Regents' Junior Faculty Fellowship from UCSB.

\renewcommand\refname{Reference}

\bibliography{prb-yll}

\end{document}